\documentclass[referee]{raa}            

\usepackage{graphicx,times}             

\begin{document}

   \title{Supernova $\beta^-$ decay of nuclides $^{53}$Fe, $^{54}$Fe, $^{55}$Fe, and $^{56}$Fe in strongly screened plasma$^*$
 \footnotetext{$^{*}supported~~by ~~the ~~National~~ Natural~~ Science~~ Foundation~~ of~~ China~~ under~~
grants~~11565020~~ and~~the~~Natural\\ Science ~~Foundation ~~of~~
Hainan~~ province ~~under~~ grant~~ 114012.$ } }

   \volnopage{Vol.0 (201x) No.0, 000--000}      
   \setcounter{page}{1}          

   \author{Jing-Jing Liu\inst{1}, and Dong-Mei Liu\inst{1}}

   \institute{College of Electronic  and Communication Engineering, Hainan Tropical Ocean University, Sanya, 572022, China. {\it liujingjing68@126.com}\\
   }

   \date{Received~~2014 month day; accepted~~2014~~month day}

\abstract{The electron screening strong effect on the electron
energy and threshold energy of the beta decay reaction. in this
paper, we study the $\beta^-$ decay rates of some iron isotopes. The
electron screening beta decay rates increase by about two orders of
magnitude. The strong screening beta decay rates due to Q-value
correction are by more than one order of magnitude higher than those
of without Q-value correction.
 }

   \authorrunning{Jing-Jing. Liu }   
   \titlerunning{Beta decay rates of nuclide $^{53}$Fe, $^{54}$Fe, $^{55}$Fe, and $^{56}$Fe}  

   \maketitle

%
%
\section{Introduction}

Beta decay plays a key role in presupernova evolution. The cooling
rates and antineutrino energy loss are strongly affected by the
beta-decay rates. Some authors (e.g., Fuller et al. 1980;
Aufderheide et al. 1990, 1994; Langanke et al. 1998, Liu et al,
2012, 2013a, 2013b, 2013c, 2013d, 2013e, 2013f, 2014, 2015, 2016a,
2016b) did a lot on the beta decay and electron capture. However,
the effect of SES on weak interaction are not included.

According to the linear response theory model (LRTM) and shell model
Fermi theory, we studied the SES beta decay rates of nuclides
$^{53}$Fe, $^{54}$Fe, $^{55}$Fe, and $^{56}$Fe in astrophysical
environments, which are very important for numerical simulation of
supernova explosions (e.g., Fuller et al. 1982; Aufderheide et al.
1990, 1993, 1994; Langanke et al. 2003; Domingo-Pardo et al. 2009).
The article is organized as follows. In section 2, we studied the
beta-decay rates by including and neglecting SES effect. In section
3, the results and discussions are presented. The conclusions are
given in section 4."

\section{Study the $\beta^-$ decay}

\subsection{The beta decay in no SES}

In no SES, the $\beta^-$ decay rates  is given by (Fuller et al.
1980; Aufderheide et al. 1990, 1994; Liu 2016a).
\begin{equation}
   \lambda_{bd}^0=\ln2 \sum\frac{(2J_i+1)e^{\frac{-E_i}{k_{B}T}}}{G(Z, A, T)} \sum_{f}
  \frac{\psi(\rho, T, Y_e, Q_{ij})}{ft_{ij}},
\label{eq:1}
\end{equation}
where $J_i$ is the spin, and $E_i$ is excitation energies of the
parent states. \textbf{$k_{B}$} is the Boltzmann constant. $ft_{ij}$
is the comparative half-life connecting states of $i$ and $j$,
$Q_{ij}$ is the nuclear energy difference between the states of $i$
and $j$. $Q_{00}=M_{p}c^2-M_{d}c^2$, $M_p$ and $M_d$ are the masses
of the parent nucleus and the daughter nucleus, respectively, $E_i$
and $E_j$, are the excitation energies of the $i$ th and $j$ th
nuclear state. $G(Z, A, T)$ is the nuclear partition function.

The  phase space integral $\psi(\rho, T, Y_e, Q_{ij})$ for the
$\beta^-$ decay is given by
\begin{equation}
\psi(\rho, T, Y_e,
Q_{ij})=\frac{c^3}{(m_ec^2)^5}\int_1^{Q_{ij}}d\varepsilon_e
\varepsilon_e(\varepsilon_e^2-1)^{1/2}(Q_{ij}-\varepsilon_e)^2\frac{F(Z+1,
\varepsilon_e)}{1+\exp[(U_F-\varepsilon_e)/k_{B}T]},
 \label{eq.2}
\end{equation}
where $p, m_e, U_F$ and $\varepsilon_e$ are the electron momentum,
mass, chemical potential and energy, respectively. $F(Z+1,
\varepsilon_e)$ is the Coulomb wave correction.

In no SES, a reasonable approximation for the electron chemical
potential takes the form (e.g., Bludman et al. 1978)
\begin{equation}
U_F=1.11(\rho_7Y_e)^{1/3}[1+(\frac{\pi}{1.11})\frac{(k_BT)^2}{(\rho_7Y_e)^{2/3}}]^{-1/3}~~
\rm{MeV}
 \label{eq.3}
\end{equation}

According to discussions from Zhou \& Li et al. (2017), the
half-life $ft_{ij}$ has the labels
\begin{equation}
\ln(ft_{ij})=a_1+(\alpha^2Z^2-5+a_2\frac{N-Z}{A})\ln(Q_{if}-a_3\delta)+(a_4\alpha^2Z^2)+\frac{1}{3}\alpha^2Z^2\ln(A)-\alpha
Z\pi+S(N,Z),
 \label{eq.4}
\end{equation}
where $\alpha$ is the fine structrue constant with $1/137$. The
correction factor $S(N,Z)$ will take the labels (e.g., Zhou \& Li et
al. 2017)
\begin{eqnarray}
S(N,Z)=&a_5\exp(-((N-28)^2+(N-20)^2)/12)+a_6\exp(-((N-50)^2+(N-38)^2)/43)\nonumber\\
&+a_7\exp(-((N-82)^2+(N-50)^2)/13)+a_8\exp(-((N-82)^2+(N-58)^2)/24)\nonumber\\
&+a_9\exp(-((N-110)^2+(N-70)^2)/244),
 \label{eq.5}
\end{eqnarray}
where ${a_i(i=,2,3,...9)}={11.09, 1.07, -0.935, -5.398, 3.016,
3.879, 1.322, 6.030, 1.669}$ in Eqs.(4-5). In Eq.(4), the shell and
pairing effect on the nuclear matrix elements, which reflect the
main information of the nuclear structure. The factor $\delta$ is
well described by $\delta=(-1)^N+(-1)^Z$ (zhou \& Li et al. 2017).

The Fermi and the Gamow-Teller matrix elements for $\beta^-$ decay
is given by (e.g., Aufderheide et al. 1994)
\begin{eqnarray}
|M_{F}(fi)|^2&=&|\langle\omega_f^D\mid\sum_n(\tau_{\pm1})|\omega_i^P\rangle|^2/(2J_i+1)
\nonumber\\
 &=&|\langle j_p||(\tau_{\pm1})||j_n\rangle|^2\frac{N_n}{(2j_n+1)(2J_i+1)}(1-\frac{N_p}{2j_p+1}),
 \label{eq.6}
\end{eqnarray}

\begin{eqnarray}
|M_{GT}(fi)|^2&=&|\langle\omega_f^D\mid\sum_n\sigma_n(\tau_{\pm1})|\omega_i^P\rangle|^2/(2J_i+1)
\nonumber\\
 &=&|\langle j_p||\sigma_n(\tau_{\pm1})||j_n\rangle|^2\frac{N_n}{(2j_n+1)(2J_i+1)}(1-\frac{N_p}{2j_p+1}),
 \label{eq.7}
\end{eqnarray}
where $|\omega_i^P\rangle$ is the initial parent state and
$|\omega_f^D\rangle$ is the final daughter state. $N_n$, and $N_p$
are the numbers of neutrons and protons within the $j_n$, and $j_p$
shell, respectively.

The Shell Model Monte Carlo (SMMC) method is used to calculate the
total amount of GT strength $S_{GT^-}$ and the response function
$R_A(\tau)$ of an operator $\hat{A}$ at an imaginary-time $\tau$.
$R_A(\tau)$ is given by (e.g., Langanke et al. 1998; Langanke et al.
2003)
\begin{small}
\begin{equation}
R_A(\tau)=\frac{\sum_{if}(2J_i+1)e^{-\beta E_i}e^{-\tau
(E_f-E_i)}|\langle f|\hat{A}|i\rangle|^2}{\sum_i (2J_i+1)e^{-\beta
E_i}}, \label{eq:9}
\end{equation}
\end{small}
where $E_i$ and $E_f$ are energies corresponding to the final states
$|i\rangle$ and $|f\rangle$. The total strength for the operator is
given by $R(\tau=0)$. The strength distribution is given by
\begin{small}
\begin{eqnarray}
\ S_{GT^+}(E) &=& \frac{\sum_{if}\delta (E-E_f+E_i)(2J_i+1)e^{-\beta
E_i}|\langle f|\hat{A}|i\rangle|^2}{\sum_i (2J_i+1)e^{-\beta E_i}} \nonumber\\
&=& S_{A}(E),
 \label{eq:10}
\end{eqnarray}
\end{small}
which is related to $R_A(\tau)$ by a Laplace Transform,
$R_A(\tau)=\int_{-\infty}^{\infty}S_A(E)e^{-\tau E}dE$. Note that
here $E$ is the energy transfer within the parent nucleus, and that
the strength distribution $S_{GT^+}(E)$ has units of $\rm
{MeV^{-1}}$ and $\beta=1/T_N$, $T_N$ is the nuclear temperature.

\subsection{The beta decay in SES}

Electron screening for nuclear reactions in astrophysical
environments plays an unexpected and important role in enhancing
reaction cross sections. In our previous works (e.g., Liu. 2013d,
2016c, 2017a, 2017b), we discussed this interesting problem. Based
on the linear response theory model (hereafter LRTM), Itoh et
al.(2002) also studied the influence of the screening potential on
the weak interaction. The electron is strongly degenerate in our
considerable regime of the density-temperature, which is given by
\begin{small}
\begin{equation}
 T\ll T_F=5.930\times10^9\{[1+1.018(\frac{Z}{A})^{2/3}(10\rho_7)^{2/3}]^{1/2}-1\},
 \label{eq.8}
\end{equation}
\end{small}
here the electron Fermi temperature and the density are $T_{\rm{F}}$
and $\rho_7$ (in units of $10^7\rm{g/cm^3}$), respectively.

Jancovici et al. (1962) studied the static longitudinal dielectric
function for relativistic degenerate electron liquid. The electron
potential energy in SES is given by
\begin{equation}
V(r)=-\frac{Ze^2(2k_{\rm{F}})}{2k_{\rm{F}}r}\frac{2}{\pi}\int_0^\infty
\frac{\rm{sin}[(2k_{\rm{F}}r)]q}{q\epsilon(q,0)}dq,
 \label{eq.9}
\end{equation}
where $\epsilon(q,0)$ is Jancovici¡¯s static longitudinal dielectric
function and $k_{\rm{F}}$ is the electron Fermi wave-number.

For relativistic degenerate electrons and based on LRTM, the
screening potential is calculated as
\begin{equation}
D=7.525\times10^{-3}Z(\frac{10z\rho_7}{A})^{\frac{1}{3}}J(r_s,R)
(\rm{MeV})
\label{eq.10}
\end{equation}
where the parameters $J(r_s,R)$, $r_s$ and $R$ are discussed by Itoh
et al.(2002)in detail. The Eqs. (12, 14) are satisfied for
$10^{-5}\leq r_s \leq 10^{-1}, 0\leq R\leq 50$, which fulfill in the
pre-supernova environment.

When we take account into the influence of SES, the beta decay
Q-value changes by (Fuller et al(1982))
\begin{equation}
\Delta Q\approx 2.940\times10^{-5}Z^{2/3}(\rho
Y_e)^{1/3}~~~\rm{MeV}.
\label{eq.11}
\end{equation}
Thus, the Q-value of beta decay changes from $Q_{if}$ to
$Q_{if}'=Q_{if}-\Delta Q$.

We can not neglect its influence at high density when electron is
strongly screened due to the screening energy is so high. The
electron screening make electron energy increase from
$\varepsilon_e$ to $\varepsilon^{s}_e=\varepsilon_e+D$ beta decay.
The screening also decreases the threshold energy from $Q_{if}$ to
$Q_{if}^s(\rm{I})=Q_{if}+D$, and
$Q_{if}^s(\rm{II})=Q_{if}'+D=Q_{if}-\Delta Q+D$ corresponding to the
SES model(I) and model(II).  The SES model(I) and model(II) are
corresponding to the case without and with the correction of
Q-value. So the phase space integral $\psi^s(\rho, T, Y_e, Q_{ij})$
replaces $\psi(\rho, T, Y_e, Q_{ij})$ in Eq.(2) for the SES beta
decay rates, and is calculated as
\begin{equation}
\psi^s(\rho, T, Y_e,
Q_{ij}^s)=\frac{c^3}{(m_ec^2)^5}\int_{1+D}^{Q_{ij}^s}d\varepsilon_e^s
\varepsilon_e^s((\varepsilon^s_e)^2-1)^{1/2}(Q^s_{ij}-\varepsilon_e^s)^2\frac{F(Z+1,
\varepsilon_e^s)}{1+\exp[(U_F-\varepsilon_e^s)/k_{B}T]},
 \label{eq.12}
\end{equation}

Therefore, according to Eq.(1), the beta decay rate in SES is given
by
\begin{equation}
   \lambda_{bd}^s=\ln2 \sum\frac{(2J_i+1)e^{\frac{-E_i}{k_{B}T}}}{G(Z, A, T)} \sum_{f}
  \frac{\psi^s(\rho, T, Y_e, Q^s_{ij})}{ft^s_{ij}},
\label{eq:13}
\end{equation}
where the half-life $ft^s_{ij}$ is given by
\begin{equation}
\ln(ft^s_{ij})=a_1+(\alpha^2Z^2-5+a_2\frac{N-Z}{A})\ln(Q^s_{if}-a_3\delta)+(a_4\alpha^2Z^2)+\frac{1}{3}\alpha^2Z^2\ln(A)-\alpha
Z\pi+S(N,Z),
 \label{eq.14}
\end{equation}

We compare the results ($\lambda_{bd}^s$) in SES with those of the
rates ($\lambda_{bd}^0$) without SES by defining an enhancement
factors $C$, which is given by
\begin{equation}
C=\frac{\lambda_{bd}^s}{\lambda_{bd}^0}.
 \label{15}
 \end{equation}

\section{Numerical results and discussions}

Based on proton-neutron quasi particle RPA model, Nabi(2010)
investigated the beta decay rates in supernova. Under the same
conditions, FFN (Fuller et al.1982) and Aufderheide et al. (1990,
1994) also discussed the beta rates. Their studies show that the
beta decay rates play an important role in the core collapse
calculations and evolution. However, they neglected the effect of
SES on beta decay. Here based on LRTM, we discuss the beta decay for
SES models (I) and (II). The model (I) and (II) are corresponding to
the case without and with correction and correction of Q-value.
Figure 1 and 2 presents the influences of density on beta decay
rates of some iron group isotopes for the two SES models. The no SES
and SES rates corresponded to solid and dotted line are compared. We
detailed the GT transition contribution for beta decay according to
SMMC method. For a given temperature, we find that the beta decay
rates decrease by more than six orders of magnitude as the density
increases. The strong screening rates always higher than those of no
SES. For example, at $\rho_7=5000, T_9=7.79,Y_e=0.45$ the rates for
$^{53}$Fe are $1.716\times10^{-17}$ and $4.065\times10^{-17}$
corresponding to those of no SES and SES for model (I) in Fig.1(b),
but are $1.129\times10^{-16}$ and $2.451\times10^{-16}$ for model
(II) in Fig.2(b). The SES beta decay rates of model (II) are by more
than one order of magnitude higher than those of model (I).

Figure 3 shows the screening enhancement factors $C$ as a function
of $\rho_7$. Due to SES, the rates may increase by about two orders
of magnitude. For instance, the screening enhancement factor $C$
increases from 11.55 to 170.8 when the density increases from $10^3$
to $10^4$ for $^{53}$Fe at $T_9=0.79,Y_e=0.48$ for model (II)in
Fig.3(a). The lower the temperature, the larger the effect of SES on
beta decay rates is. One possible cause that the SES mainly
increased the number of higher energy electrons. These electrons can
actively join in the beta decay reaction. Moreover, the SES can also
make the beta decay threshold energy greatly decrease. Thus, SES
strong encourage the beta decay reactions. One also find that the
SES enhancement factor $C$ of model (II) are higher than those of
model (I). For example, at $\rho_7=7000, T_9=0.79,Y_e=0.48$ the
screening factor $C$ for $^{53,54,55,56}$Fe is 88.09, 86.18, 70.11,
70.53 for model (I), and are 98.36, 95.89, 82.56, 84.12 for model
(II) in Fig.3(a), respectively.

Table 1 and 2 show the screening enhancement factor $C$ at
$\rho_7=1000, 10000$ for model (I), and (II). From Table 1, the
screening rates for $^{53,54,55,56}$Fe increase by a factor of
10.59, 10.43, 9.288, 9.349 at $\rho_7=1000, T_9=0.79,Y_e=0.48$ for
model (I), and by a factor of 11.56, 11.46, 10.19, 10.32 for model
(II), respectively. From Table 2, the screening rates for
$^{53,54,55,56}$Fe increase by a factor of 155.7, 150.6, 118.9,
120.6 at $\rho_7=10000, T_9=0.79,Y_e=0.48$ for model (I), and by a
factor of 170.8, 166.4, 132.2, 135.2 for model (II), respectively.
But the difference of the screening enhancement factor $C$ between
model (I), and (II) is small at the higher temperature. This is
because that the higher the temperature, the larger the electron
energy is for a given density. So the higher temperature weaken the
effect of SES on beta decay.

The beta decay rates are strong depended on the decay Q-value. The
higher the energy of outgoing electron, the larger the rates become
when the electron energy is more than the threshold energy. When we
take account the Q-value correction in model (II), according to
Eq.(4), the half-life will increase as the the Q-value increase. The
nuclear binding energy increases because of interactions with the
dense electron gas in the plasma. The beta decay Q-value
($Q_{\rm{if}}$), changes at high density due to the affect of the
charge dependence of this binding. Based on Eq.(11), Q-value of beta
decay decreases from $Q_{\rm{if}}$ to $Q_{\rm{if}}-\Delta Q$.  Thus,
the beta decay will increase due to correction of Q-value in model
(II) according to Eq.(1) and Eq.(4).

In supernova evolution the distributions of Gamow-Teller strength
play a key role. As examples for the excited state GT distributions
of $^{55,56}$Fe, Fig.4 presents some information about the
comparison of our results by SMMC  with those of Nabi (Nabi et al.
2010) for beta decay. We  the first two excited state distributions
are only shown. From fig. 4, one finds that our results of GT
strength distributions calculated are lower than those of Nabi. For
example, the $\rm{GT}^{-}$ distributions for $^{55}$Fe are 1.650MeV,
1.362MeV corresponding to Nabi's and ours at $E_i=3.86\rm{MeV},
E_j=7.461\rm{MeV}$, and are 0.7265MeV, 0.5865MeV for $^{56}$Fe at
$E_i=5.18\rm{MeV}, E_j=6.055\rm{MeV}$. Based the pn-QRPA theory,
Nabi et al. (2010) analyzed nuclear excitation energy distribution
by considering the particle emission processes. They calculated
Gamow-Teller strength distribution and only discussed the low
angular momentum states. By using the method of SMMC, actually we
discuss GT intensity distribution and adopt an average distribution.

Synthesizes the above analysis, the charge screening strong effects
the beta decay. The influence may be mainly come from following
several factors. First, the electron Coulomb wave function is
strongly changed by the screening potential in nuclear reactions.
Second, the energy of outgoing electrons increases greatly due to
the electron screening potential. Third, the energy of atomic nuclei
also increases because of the electron screening (i.e., increases
the single particle energy). Finally, the electron screening
effectively make the number of the higher-energy electrons increase.
So the electron energy is more than the threshold of beta decay. SES
relatively decreases the threshold needed for beta decay.

\begin{figure*}
\centering
\includegraphics[width=7cm,height=7cm]{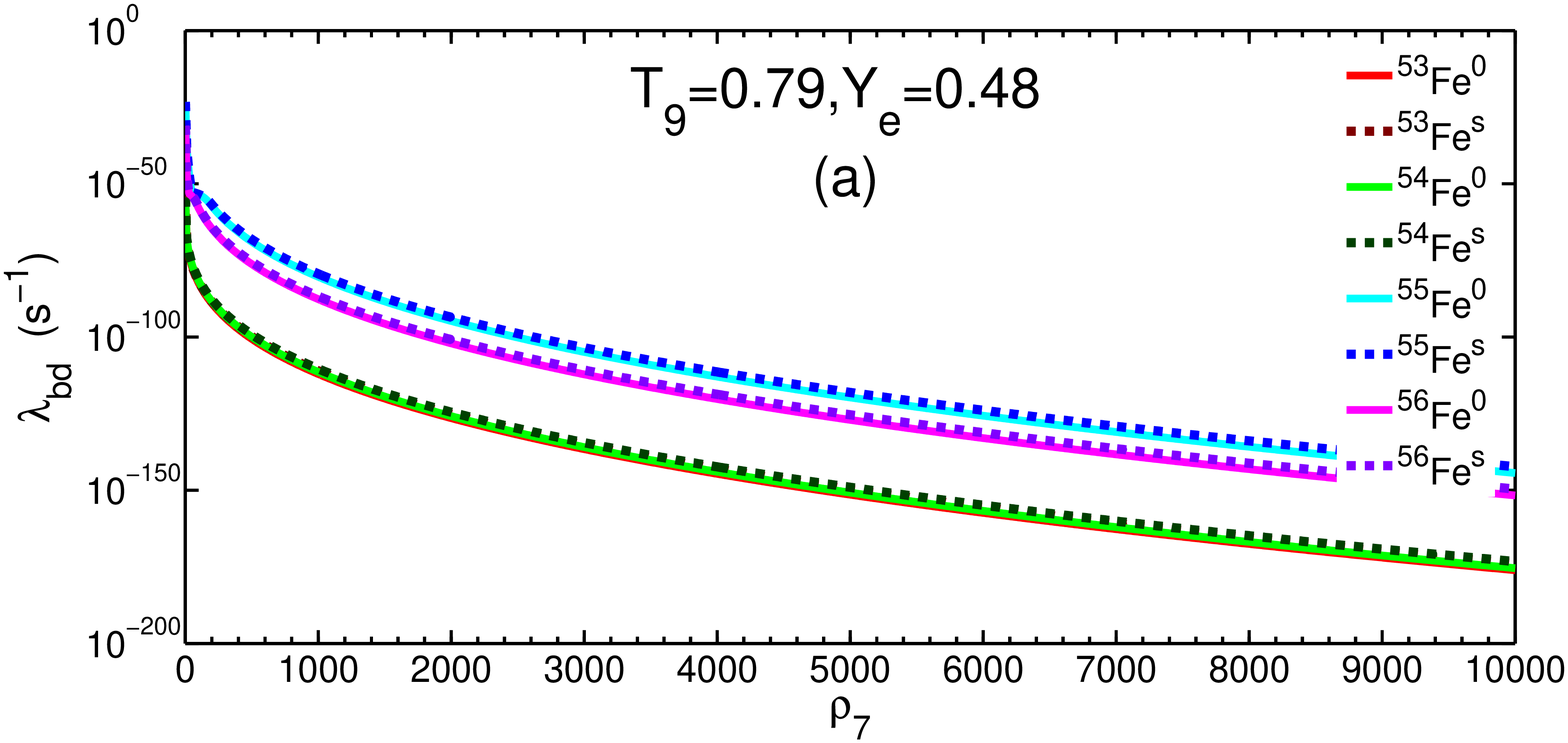}
\includegraphics[width=7cm,height=7cm]{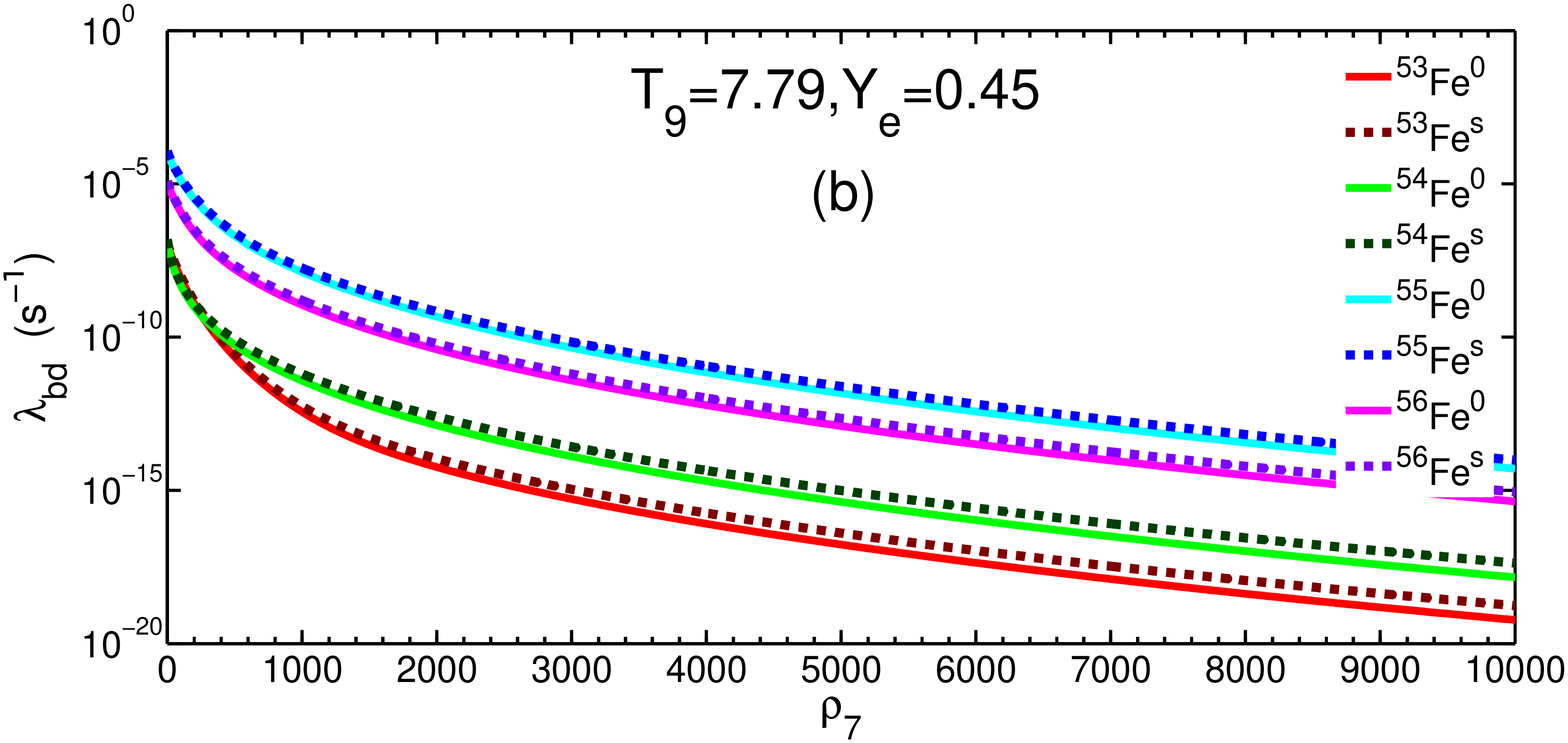}
\includegraphics[width=7cm,height=7cm]{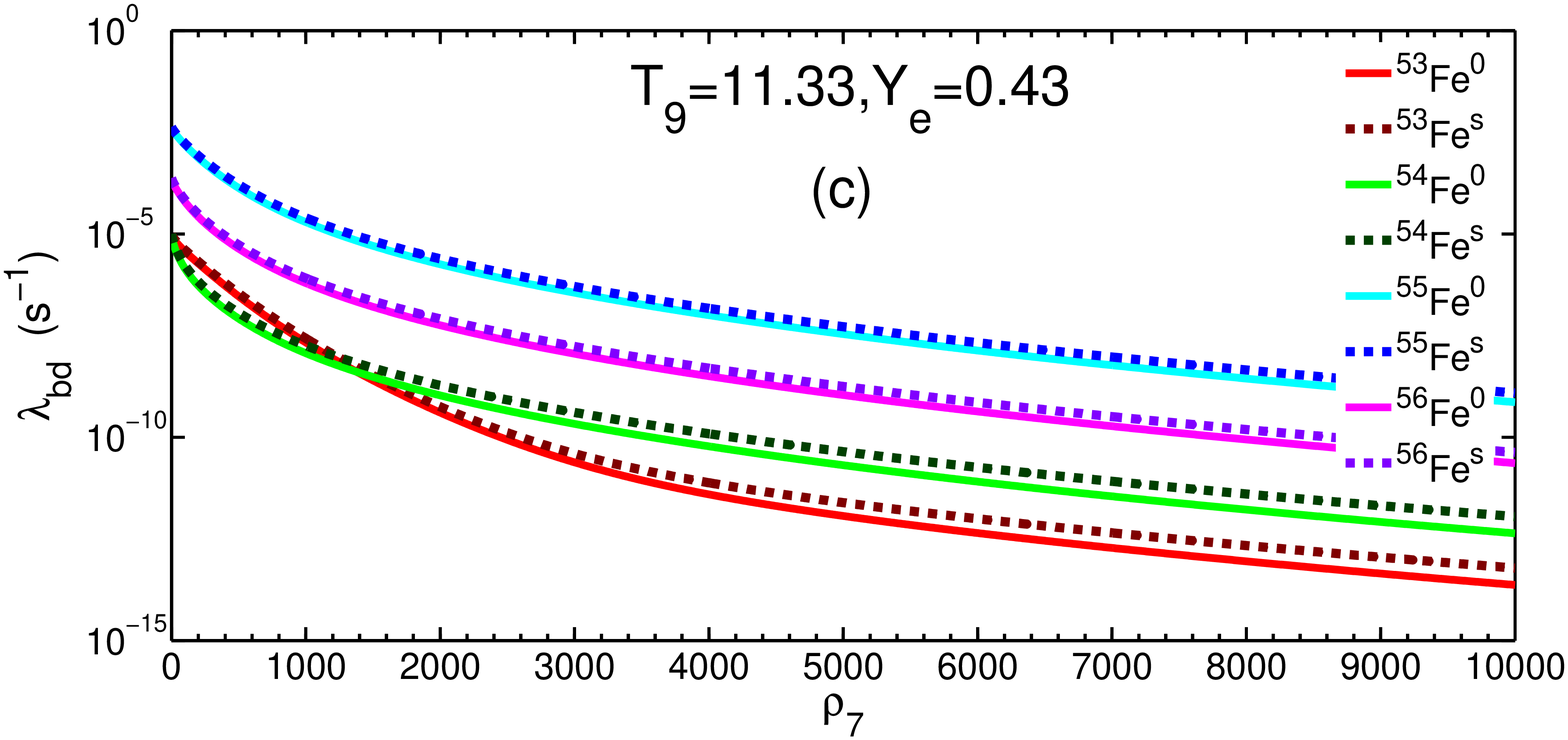}
\includegraphics[width=7cm,height=7cm]{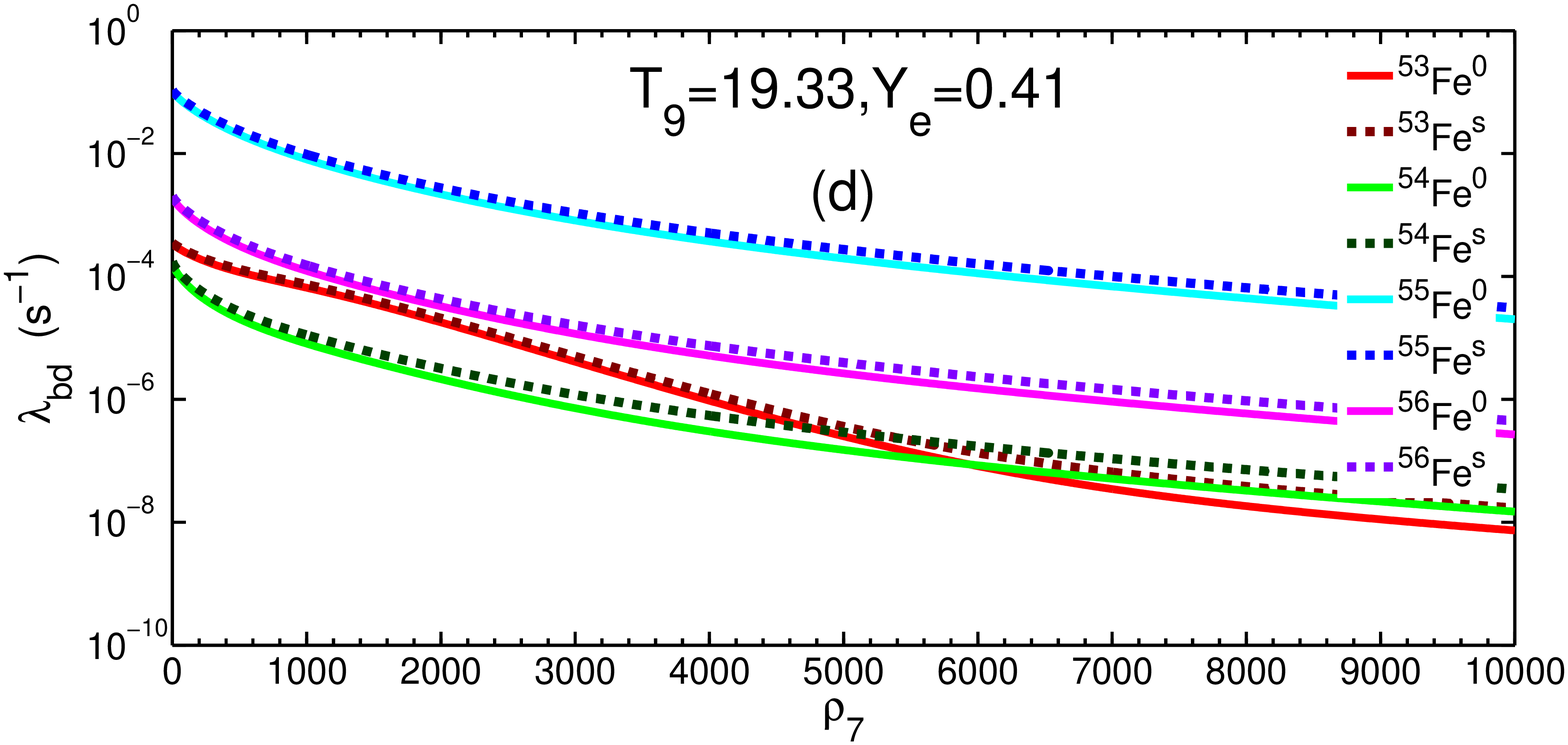}
\caption{The beta decay rates of  $^{53}$Fe, $^{54}$Fe, $^{55}$Fe,
and $^{56}$Fe as a function of electron density $\rho_7$ in and not
in SES for model (I).} \label{fig1}
\end{figure*}
\begin{figure*}
\centering
\includegraphics[width=7cm,height=7cm]{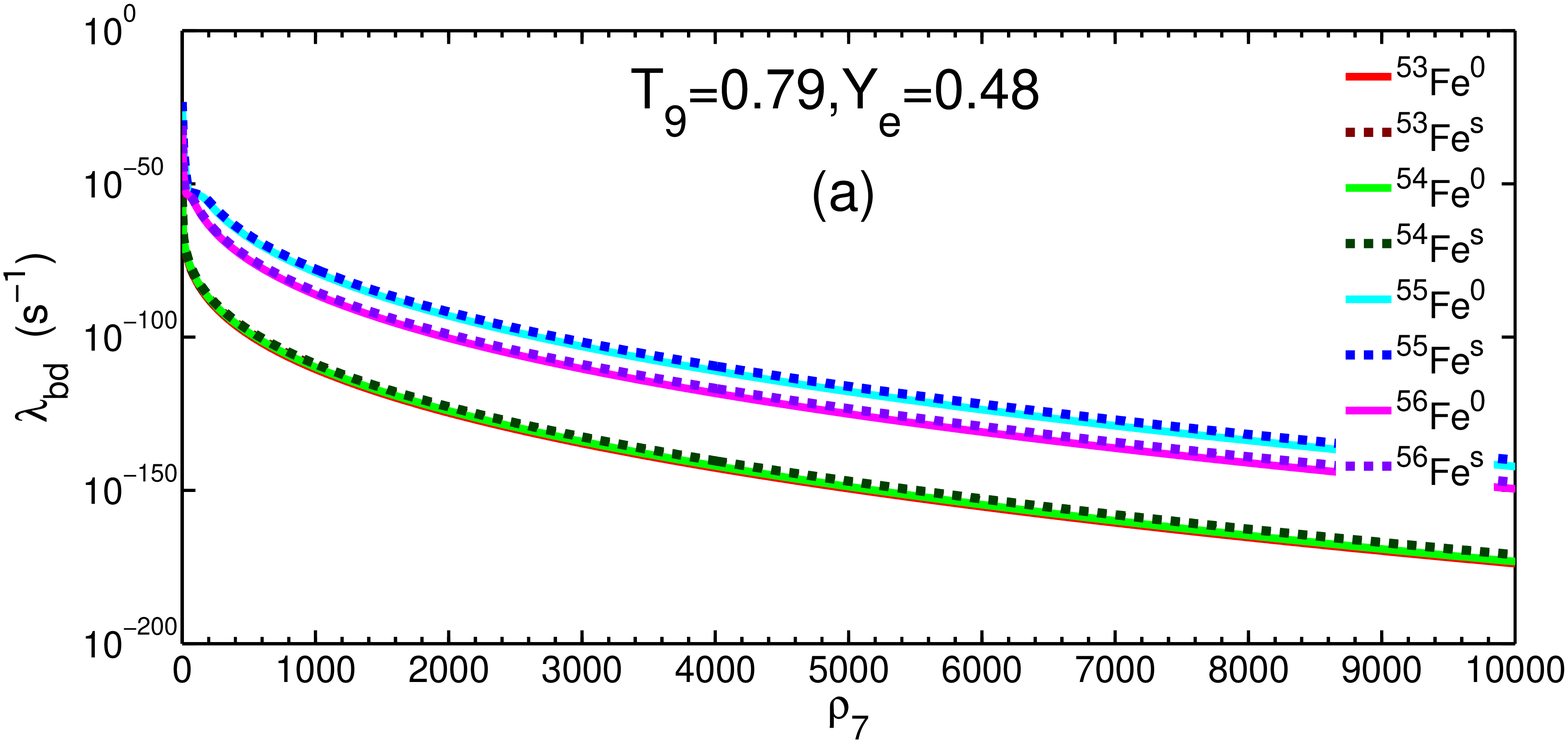}
\includegraphics[width=7cm,height=7cm]{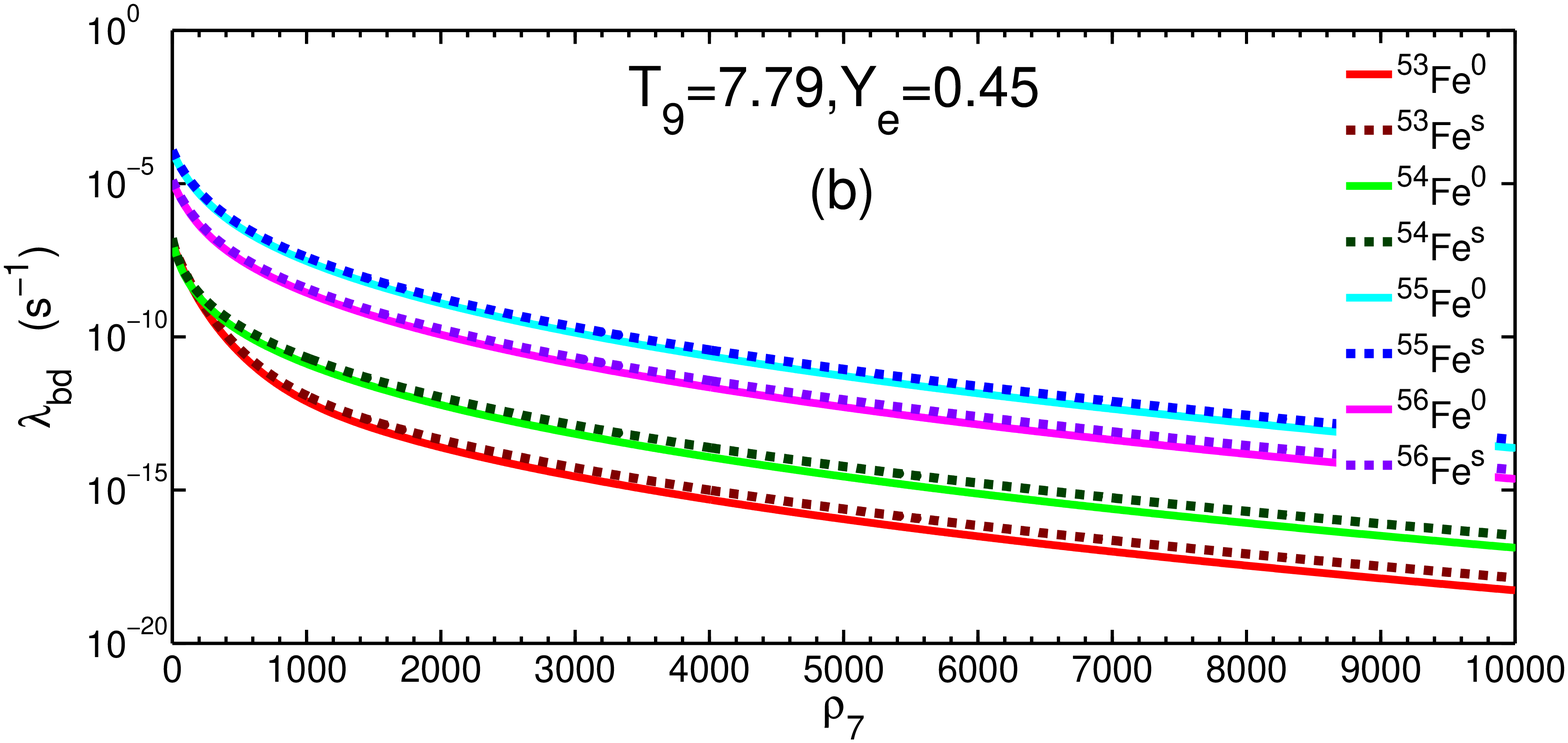}
\includegraphics[width=7cm,height=7cm]{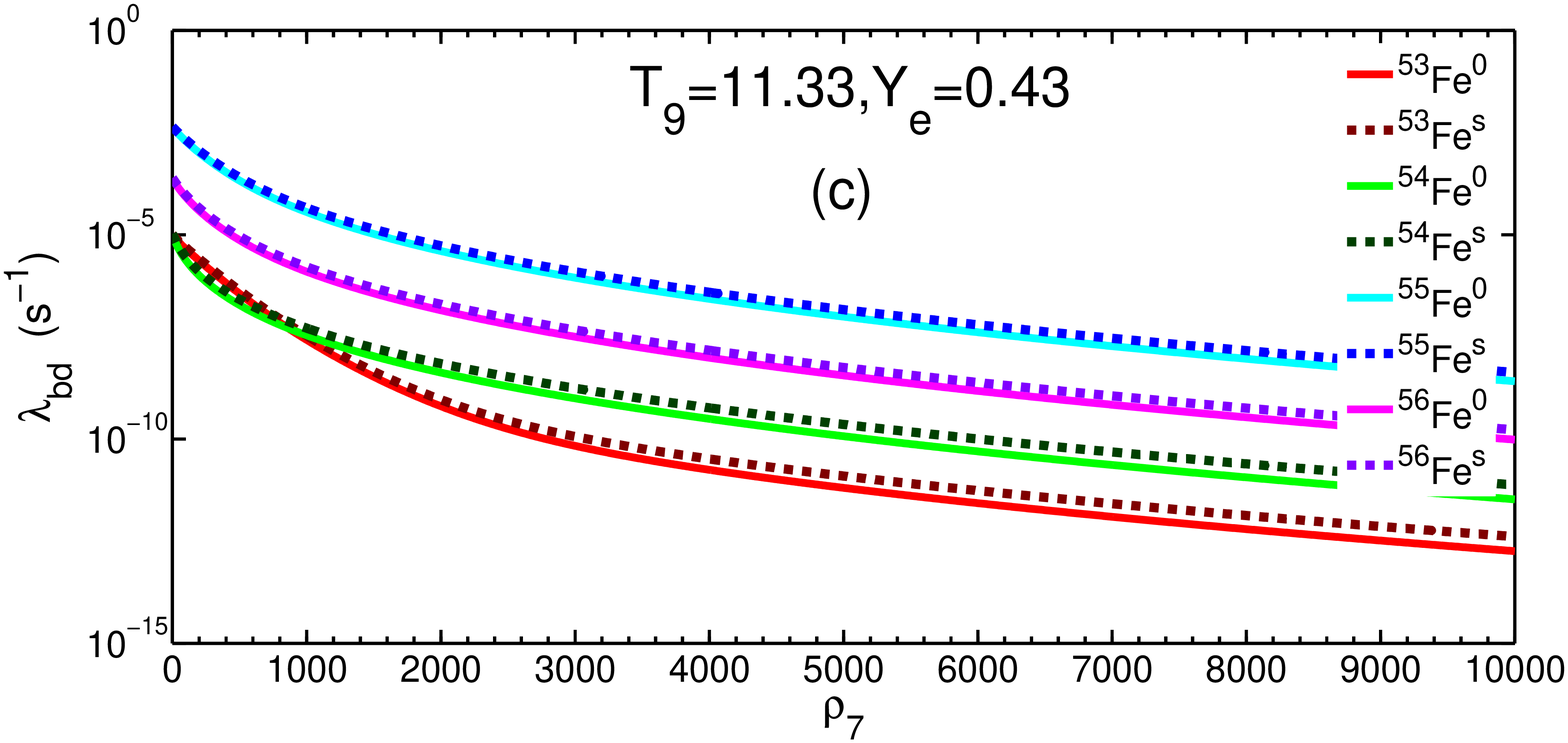}
\includegraphics[width=7cm,height=7cm]{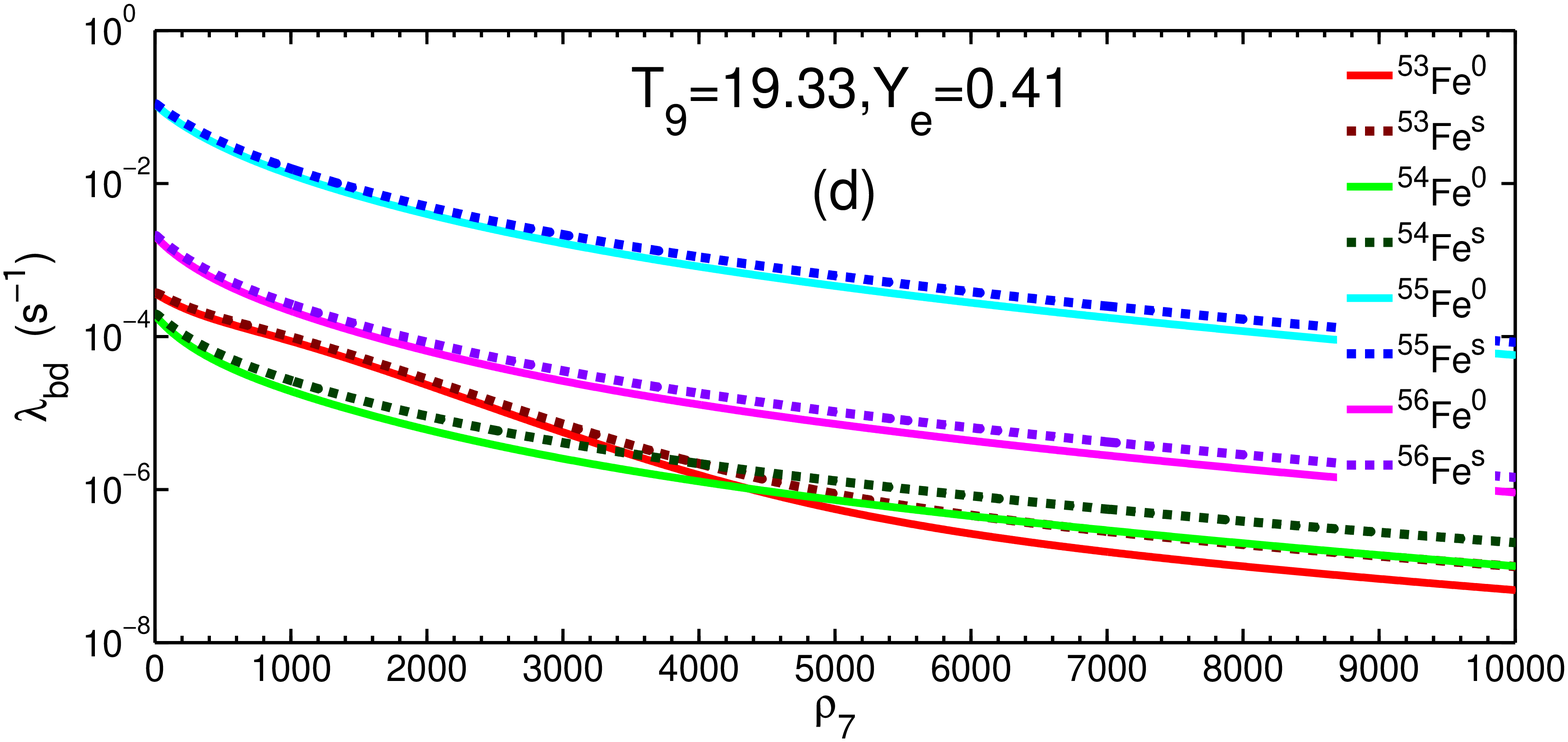}
\caption{The beta decay rates of  $^{53}$Fe, $^{54}$Fe, $^{55}$Fe,
and $^{56}$Fe as a function of electron density $\rho_7$ in and not
in SES for model (II).} \label{fig1}
\end{figure*}


\begin{figure*}
\centering
\includegraphics[width=7cm,height=7cm]{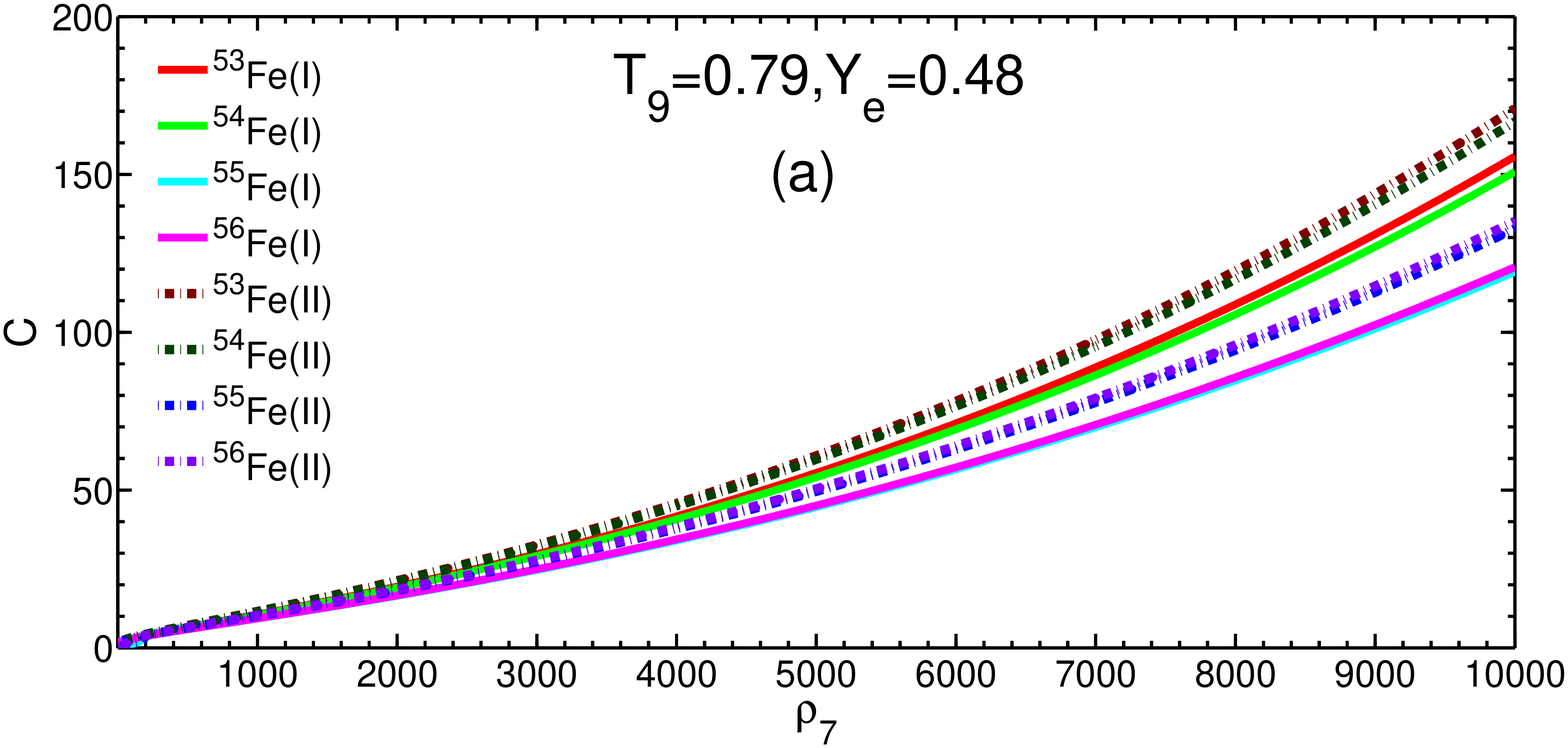}
\includegraphics[width=7cm,height=7cm]{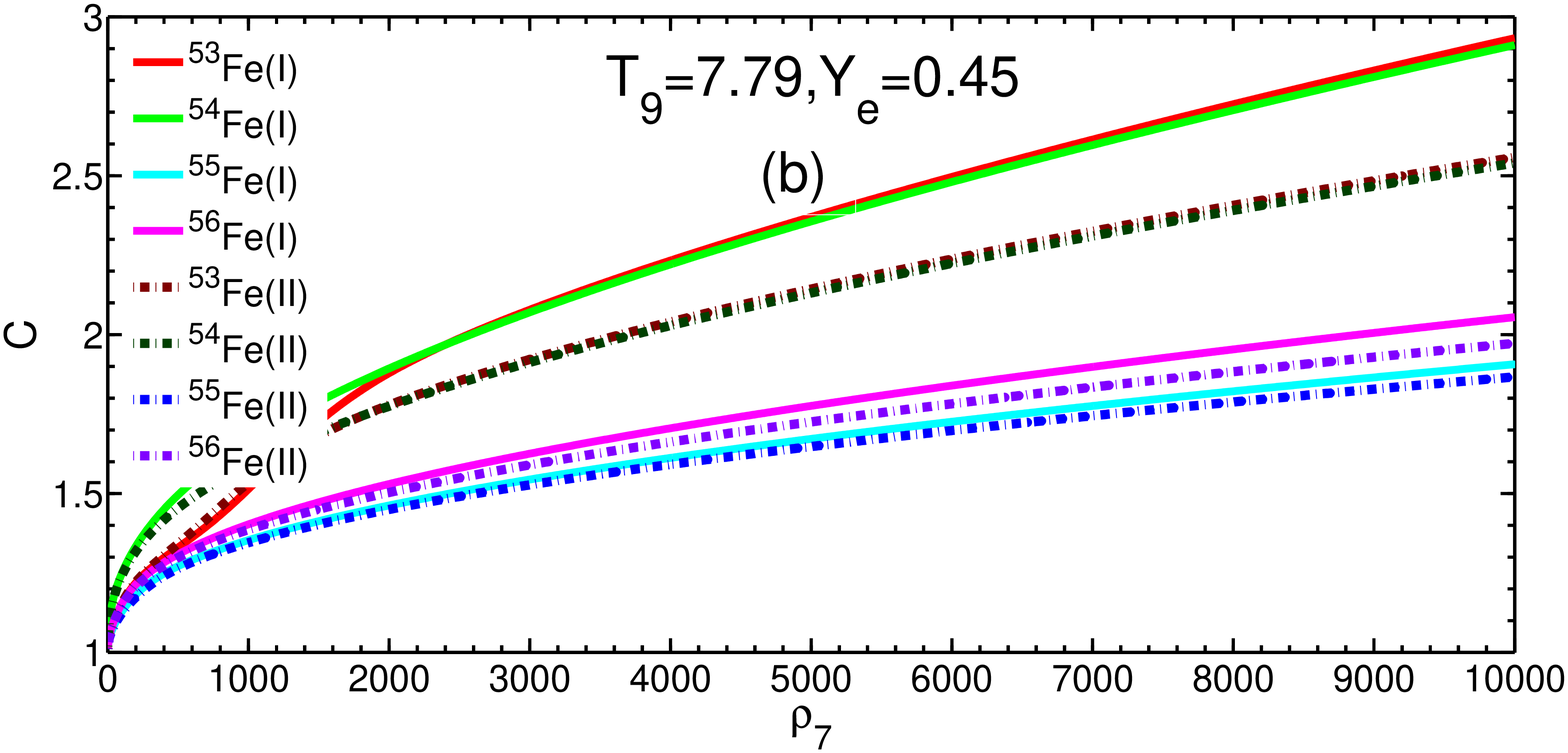}
\includegraphics[width=7cm,height=7cm]{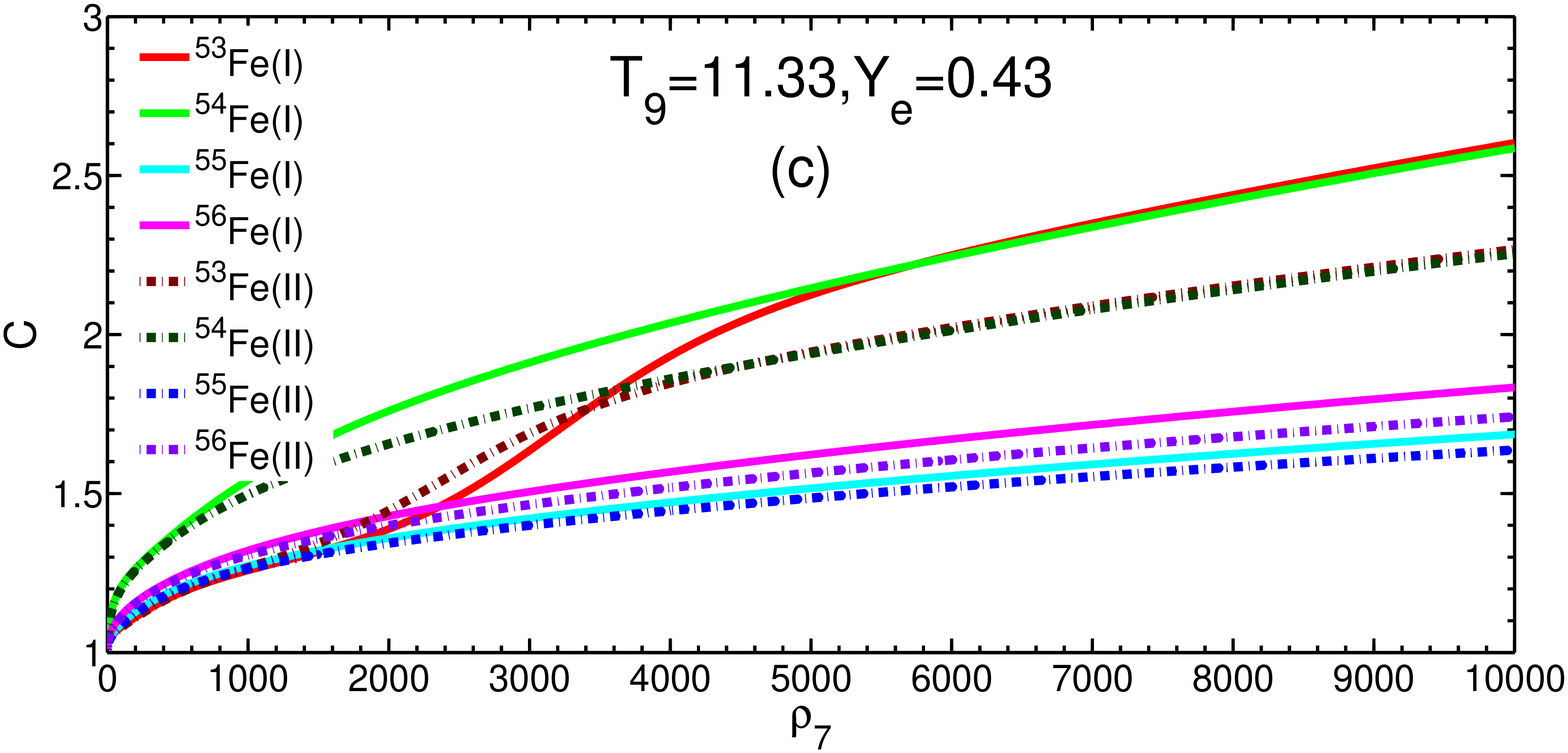}
\includegraphics[width=7cm,height=7cm]{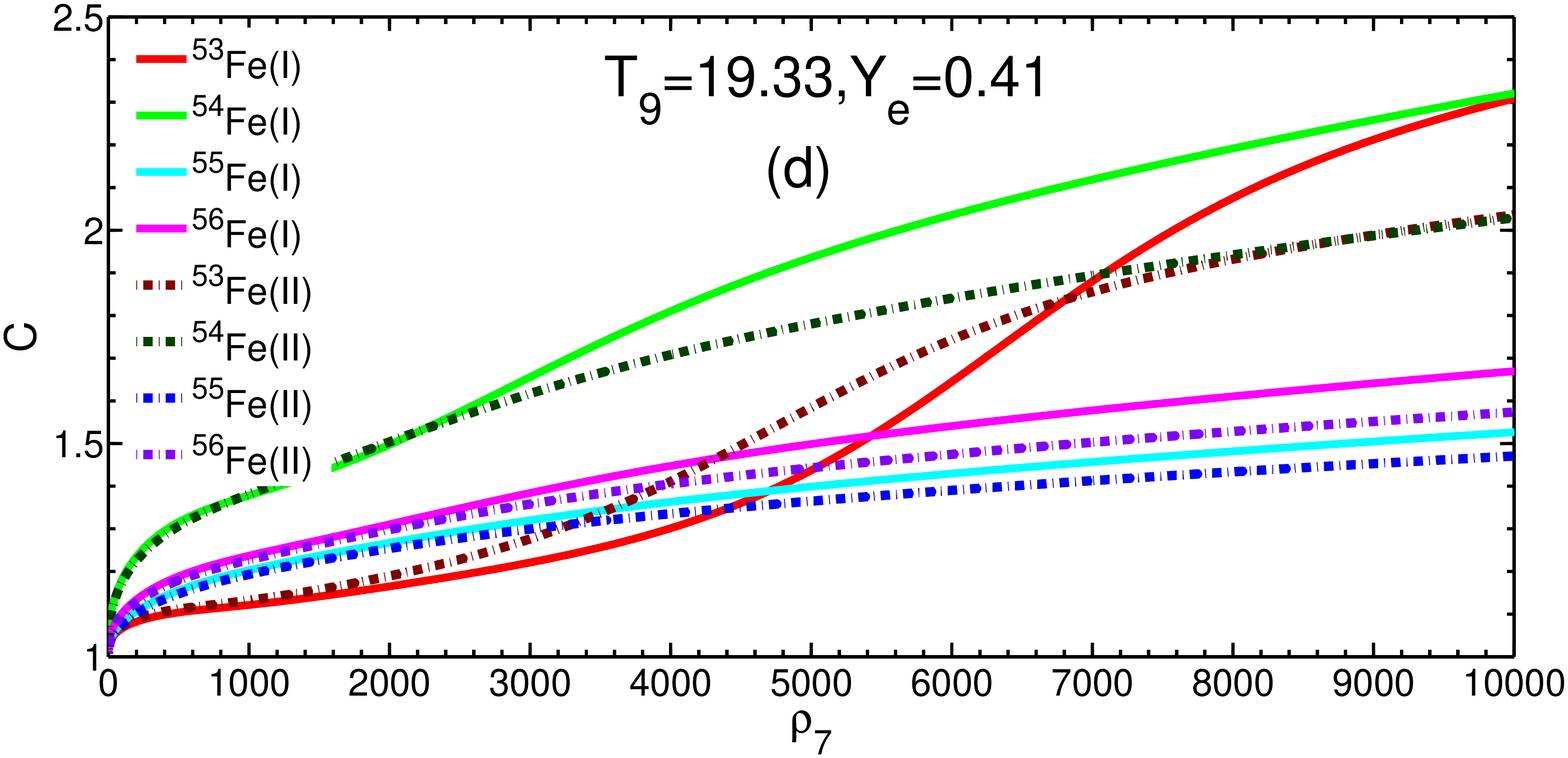}

\caption{The screening enhancement factor C for beta decay rates of
$^{53}$Fe, $^{54}$Fe, $^{55}$Fe, and $^{56}$Fe as a function of
electron density $\rho_7$ for model (I) and (II).} \label{fig2}
\end{figure*}

\begin{figure*}
\centering
\includegraphics[width=7cm,height=7cm]{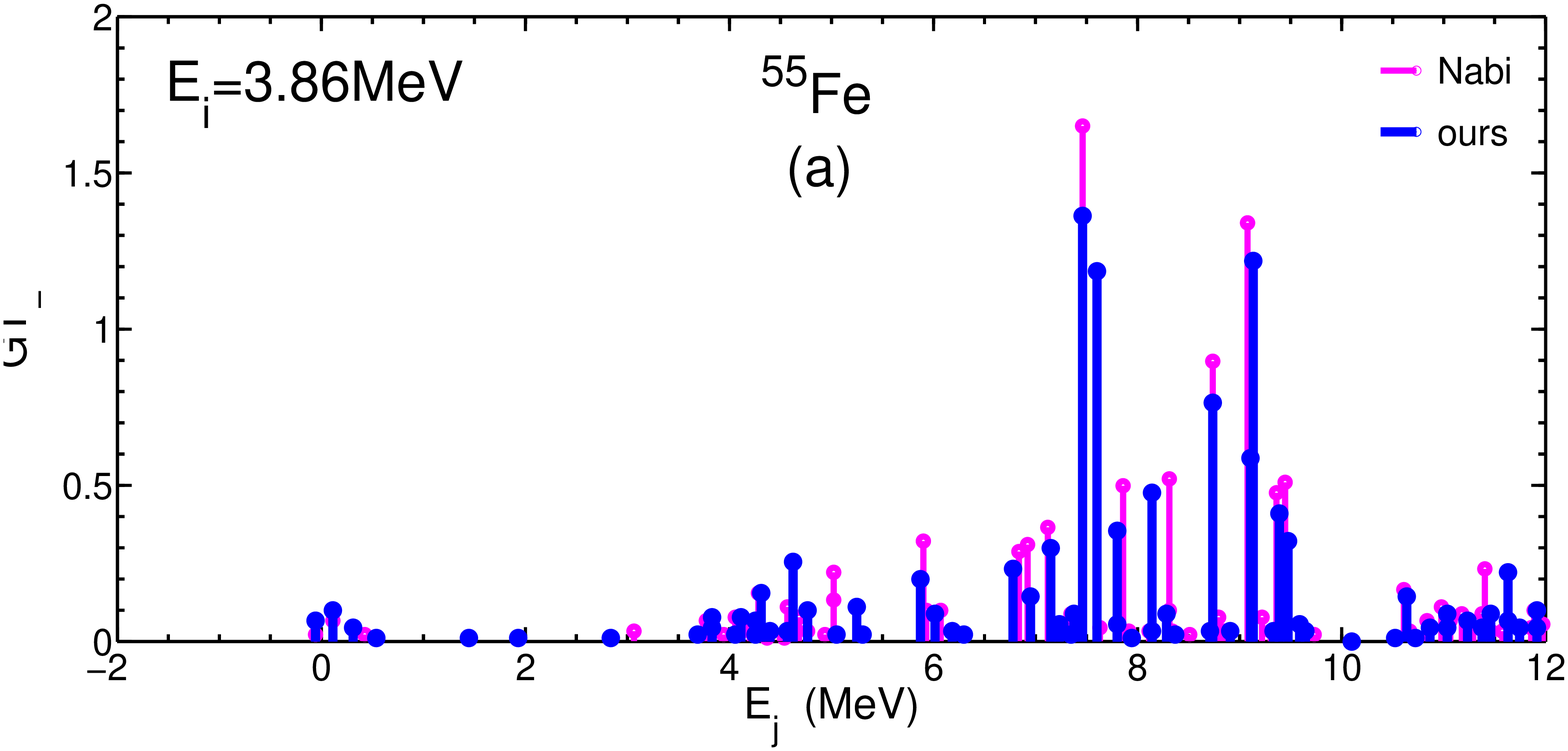}
\includegraphics[width=7cm,height=7cm]{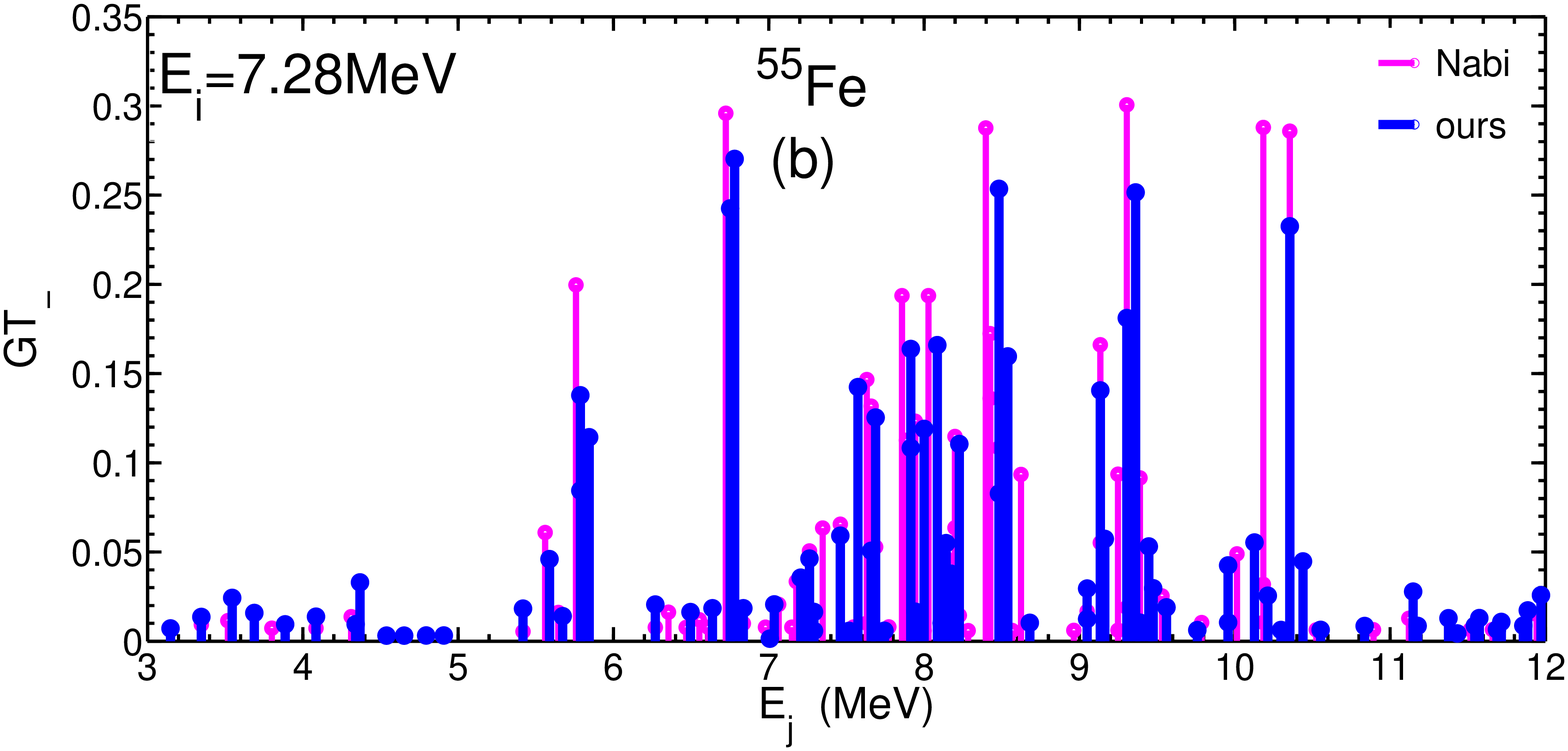}
\includegraphics[width=7cm,height=7cm]{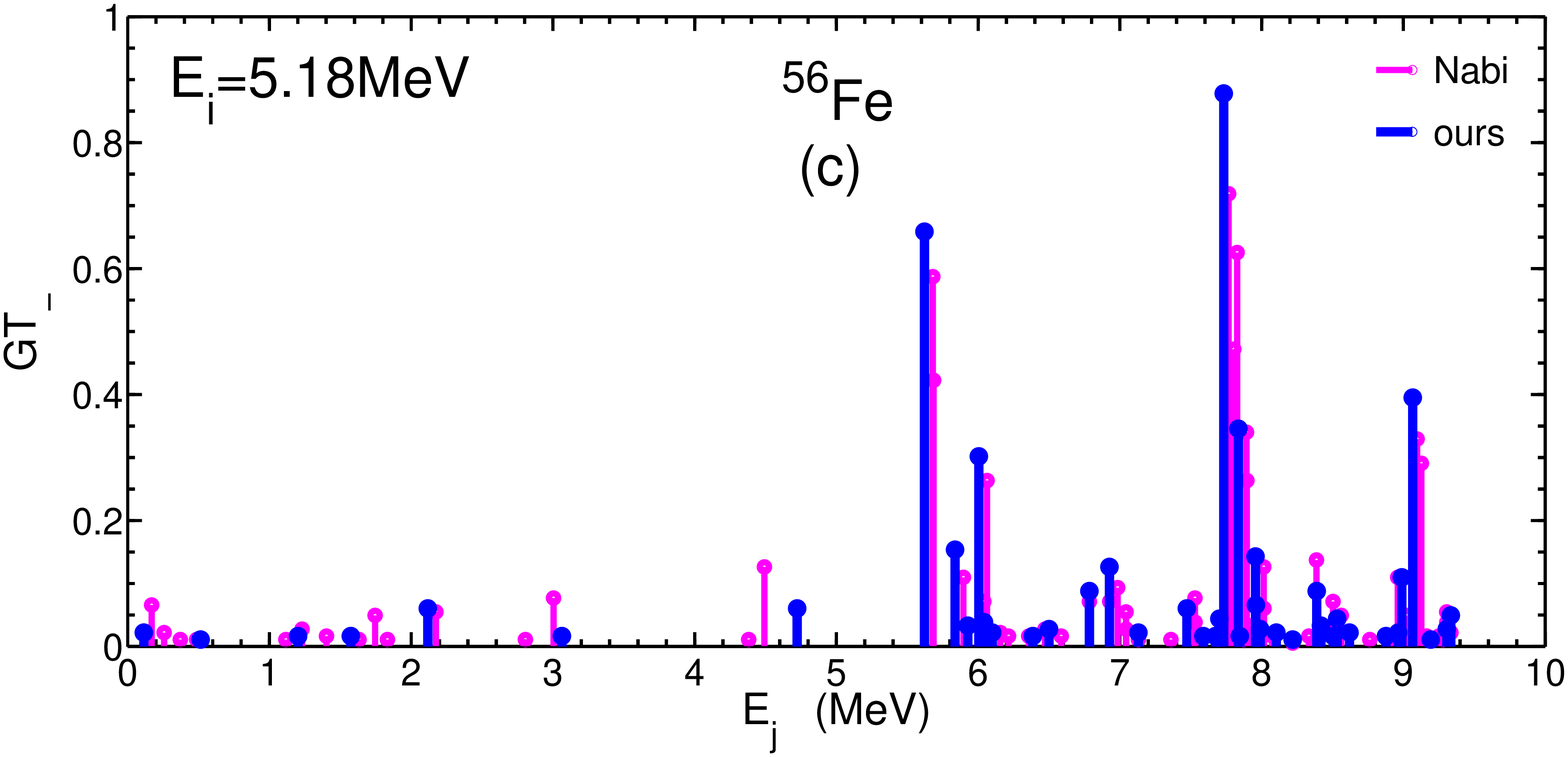}
\includegraphics[width=7cm,height=7cm]{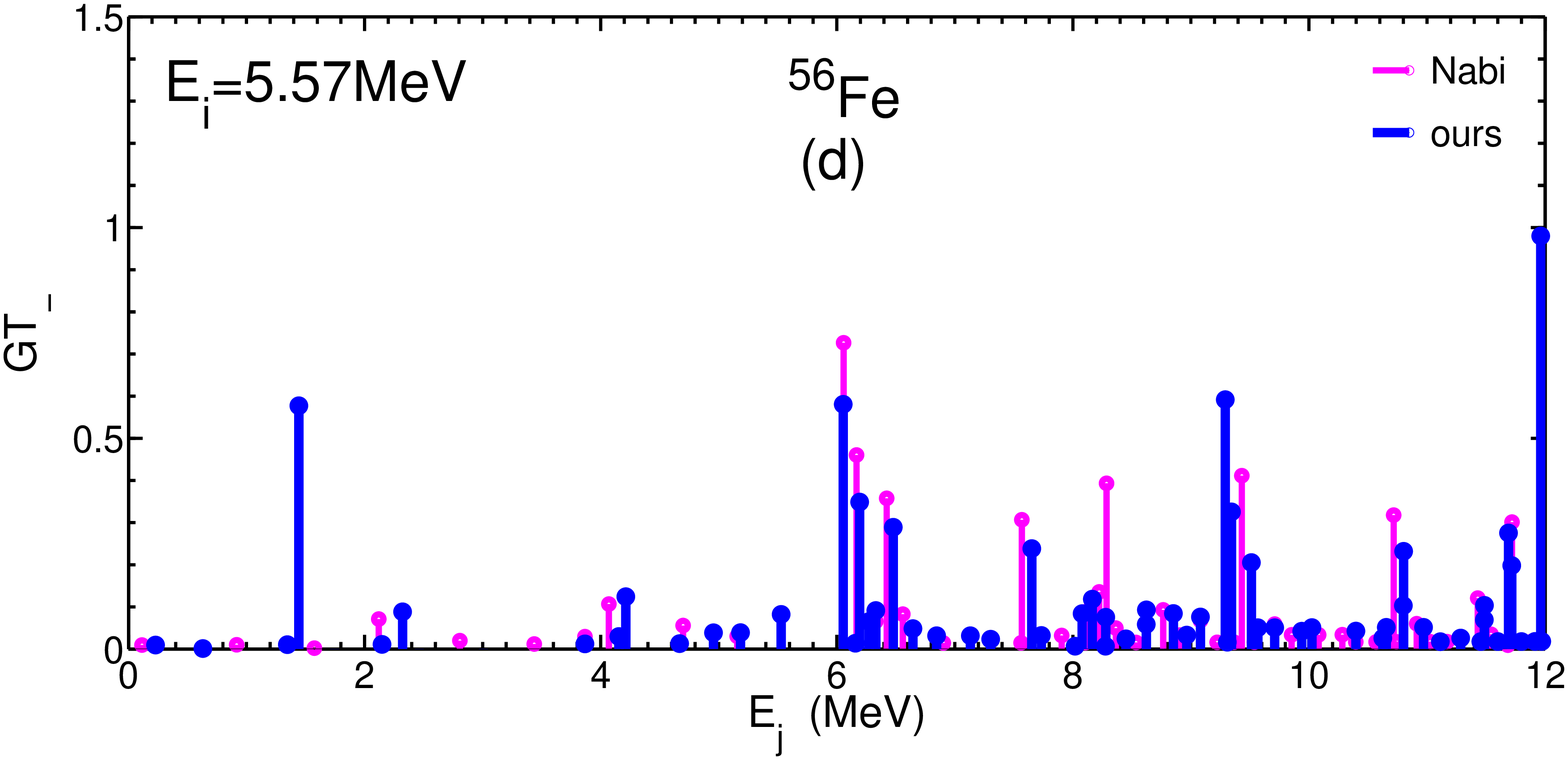}

\caption{Comparisons of the excited state Gamow-Teller strength
distributions for $^{55}$Fe, and $^{56}$Fe between ours and Nabi,
$E_i (E_j)$ represents parent (daughter) energy states.}
\label{fig3}
\end{figure*}


\begin{table*}
 \caption{The strong screening enhancement factor C for model (I) and (II) at $\rho_7=1000$ in some typical astronomical conditions.}
\centering
 \begin{minipage}{150mm}
  \begin{tabular}{@{}rrrrrrrrrrrr@{}}
  \hline
 & \multicolumn{2}{c}{$T_9=0.79, Y_e=0.48$} & &\multicolumn{2}{c}{$T_9=7.79, Y_e=0.45$}&&\multicolumn{2}{c}{$T_9=11.33, Y_e=0.43$}&&\multicolumn{2}{c}{$T_9=19.33, Y_e=0.481$}\\

\cline{2-3} \cline{5-6} \cline{8-9} \cline{11-12}\\
 nuclei &$C(\rm{I})$ & $C(\rm{II})$& & $C(\rm{I})$ & $C(\rm{II})$ & &$C(\rm{I})$ & $C(\rm{II})$& &$C(\rm{I})$ & $C(\rm{II})$  \\

 \hline
 $^{53}$Fe  &10.59 &11.56   & &1.523  &1.563   & &2.602  &2.266     & &1.121   &1.133  \\
 $^{54}$Fe  &10.43 &11.46   & &1.686  &1.611   & &2.584  &2.252     & &1.376   &1.377      \\
 $^{55}$Fe  &9.288 &10.19   & &1.347  &1.353   & &1.686  &1.636     & &1.203   &1.193   \\
 $^{56}$Fe  &9.349 &10.32   & &1.415  &1.399   & &1.833  &1.741     & &1.237   &1.228   \\

\hline
\end{tabular}
\end{minipage}
\end{table*}


\begin{table*}
 \caption{The strong screening enhancement factor C for model (I) and (II) at $\rho_7=10000$ in some typical astronomical conditions.}
\centering
 \begin{minipage}{150mm}
  \begin{tabular}{@{}rrrrrrrrrrrr@{}}
  \hline
 & \multicolumn{2}{c}{$T_9=0.79, Y_e=0.48$} & &\multicolumn{2}{c}{$T_9=7.79, Y_e=0.45$}&&\multicolumn{2}{c}{$T_9=11.33, Y_e=0.43$}&&\multicolumn{2}{c}{$T_9=19.33, Y_e=0.481$}\\

\cline{2-3} \cline{5-6} \cline{8-9} \cline{11-12}\\
  nuclei &$C(\rm{I})$ & $C(\rm{II})$& & $C(\rm{I})$ & $C(\rm{II})$ & &$C(\rm{I})$ & $C(\rm{II})$& &$C(\rm{I})$ & $C(\rm{II})$  \\

 \hline
 $^{53}$Fe  &155.7 &170.8   & &2.931  &2.566   & &1.258  &1.267     & &2.320   &2.034  \\
 $^{54}$Fe  &150.6 &166.7   & &2.911  &2.539   & &1.542  &1.497     & &2.332   &2.028      \\
 $^{55}$Fe  &118.9 &132.8   & &1.905  &1.866   & &1.273  &1.264     & &1.526   &1.470   \\
 $^{56}$Fe  &120.6 &135.2   & &2.055  &1.972   & &1.321  &1.306     & &1.669   &1.573   \\

\hline
\end{tabular}
\end{minipage}
\end{table*}

\section{Concluding remark}
Based on LRTM and Fermi theory, we discuss the beta decay process
for two typical SES models (i.e., model (I) and (II)). We detailed
the GT transition contribution to the beta decay according to SMMC
method. For a given temperature, the beta decay rates decrease by
more than six orders of magnitude with the increasing of the
density. The strong screening rates always higher than those of no
SES. The SES beta decay rates of model (II) are by more than one
order of magnitude higher than those of model (I). Our results show
that the beta decay rates increase by about one order of magnitude
due to SES. For instance, the screening enhancement factor $C$
increases from 11.55 to 170.8 when the density increases from $10^3$
to $10^4$ for $^{53}$Fe at $T_9=0.79,Y_e=0.48$ for model (II). The
beta decay rates and the antineutrino energy loss are quite relevant
for numerical simulations of stellar thermal evolution. Our results
may be helpful to the future studies of burst mechanism of
supernova, and cooling numerical simulation.

\begin{acknowledgements}
This work was supported in part by the National Natural Science
Foundation of China under grants 11565020, and the Counterpart
Foundation of Sanya under grant 2016PT43, the Special Foundation of
Science and Technology Cooperation for Advanced Academy and Regional
of Sanya under grant 2016YD28, the Scientific Research Starting
Foundation for 515 Talented Project of Hainan Tropical Ocean
University under grant RHDRC201701, and the Natural Science
Foundation of Hainan Province under grant 114012.
\end{acknowledgements}

\label{lastpage}


\begin{thebibliography}{99}

  \bibitem[1990]{Aufderheide90} Aufderheide, M. B., Brown, G. E., kuo, T. T. S., Stout, D. B., \&  Vogel, P., 1990, ApJ, 62, 241
  \bibitem[1994]{Aufderheide94} Aufderheide, M. B., Fushikii, I., Woosely, S. E., \&  Hartmanm, D. H.,, 1994, ApJS,, 91, 389
  \bibitem[1993]{Aufderheide93} Aufderheide, M. B., Bloom, S. D., Resler, D. A., Mathews, G. J., 1993, Phys.Rev. C, 91, 389
  \bibitem[1978]{Bludman78} Bludman S.A. and van Riper K.A. , 1978, Astrophys. J., 224, 631
  \bibitem[2009]{Domingo09}Domingo-Pardo, C., Dillmann, I., Faestermann, T., et al., 2009, Conference Proceedings, 1090, 230
  \bibitem[1982]{Full82} Fuller, G. M., Fowler, W. A., \&  Newman, M. J., 1982, ApJS, 48, 279
  \bibitem[2001]{Itoh02} Itoh, N., Tomizawa, N., Tamamura, M. et al., 2002, ApJ., 579, 380
  \bibitem[1962]{Jancovici62} Jancovici, B., 1962, Nuovo Cimento, 25, 428
  \bibitem[1998]{Langanke98} Langanke, K., \&  Martinez-Pinedo, G., 1998,  Phys. Lett. B, 436, 19
  \bibitem[2003]{Langanke03}Langanke, K.; Terasaki, J.; Nowacki, F.,et al.,2003, Phys.Rev. C, 67, 044314
  \bibitem[2007]{liu07} Liu, J. J. \&  Luo, Z. Q., 2007a, ChPhL, 24, 1861
  \bibitem[2007]{liu07} Liu, J. J. \&  Luo, Z. Q., 2007b, ChPhy,. 16, 3624
  \bibitem[2007]{liu07} Liu, J. J., Luo, Z. Q., Liu, H. L. \&  Lai, X. J., 2007c, IJMPA,. 22, 3305
  \bibitem[2007]{liu07} Liu, J. J. \&  Luo, Z. Q., 2007d, ChPhy, 16 2671
  \bibitem[2008]{liu08} Liu, J. J. \&  Luo, Z. Q., 2008a, ChPhC, 32, 617
  \bibitem[2008]{liu08} Liu, J. J. \&  Luo, Z. Q., 2008b, ChPhC, 32, 108
  \bibitem[2008]{liu08} Liu, J. J. \&  Luo, Z. Q., 2008c, CoTPh, 49, 239
  \bibitem[2010]{liu10} Liu, J. J., 2010a, ChPhC, 34, 171
  \bibitem[2010]{liu10} Liu, J. J., 2010b, ChPhC, 34, 1700
  \bibitem[2011]{liu11} Liu, J. J. \&  Kang, X. P. et al., 2011, ChPhC, 35, 243
  \bibitem[2012]{Liu12} Liu, J. J., 2012, ChPhL,29,122301
  \bibitem[2013]{Liu13} Liu, J. J., 2013a, Ap\&SS,343, 117
  \bibitem[2013]{Liu13} Liu, J. J., 2013b, Ap\&SS,343, 579
  \bibitem[2013]{Liu13} Liu, J. J., 2013c, RAA,13, 99
  \bibitem[2013]{Liu13} Liu, J. J., 2013d, MNRAS., 433, 1108
  \bibitem[2013]{Liu13} Liu J. J., 2013e, RAA., 13, 945
  \bibitem[2013]{Liu13} Liu J. J., 2013f, Chin.Phys. C., 37, 085101
  \bibitem[2013]{Liu14} Liu, J. J., 2014, MNRAS., 438, 390
  \bibitem[2015]{Liu15} Liu, J. J., 2015, Ap\&SS, 357, 93
  \bibitem[2016]{Liu16} Liu, J. J., et al., 2016a, RAA, 16, 174
  \bibitem[2016]{Liu16} Liu, J. J., et al., 2016b, ApJS, 224, 29
  \bibitem[2016]{Liu16} Liu, J. J., et al., 2016c, RAA, 16, 83
  \bibitem[2017]{Liu17} Liu, J. J., et al., 2017a, RAA, 17, 107, eprint arXiv:1707.03504
  \bibitem[2017]{Liu17} Liu, J. J., et al., 2017b, ChPhC, 41, 5101
  \bibitem[2010]{Nabi10} Nabi, J-U., 2010, AdSpR, 46, 1191
  \bibitem[2017]{Zhou17} Zhou, Y., Li, Z. H., Wang, Y. B., et al., 2017, SCIENCE CHINA Physics, Mechanics \& Astronomy.,10.1007/s11433-017-9045-0

\end{thebibliography}
\end{document}